\documentclass[aps,prc,twocolumn,floatfix,showpacs,amsfonts,amssymb,amsmath]{revtex4}

\usepackage{graphicx}
\usepackage{dcolumn}

\begin{document}

%\sloppy

\title{Resonance parameters of the first 1/2$^{+}$ state in $^{9}$Be and astrophysical implications}

\author{O. \surname{Burda}}
\affiliation{Institut f\"{u}r Kernphysik, Technische Universit\"{a}t
Darmstadt, D-64289 Darmstadt, Germany}

\author{P. \surname{von Neumann-Cosel}}
%\altaffiliation{Email address: vnc@ikp.tu-darmstadt.de}
\email{vnc@ikp.tu-darmstadt.de} \affiliation{Institut f\"{u}r
Kernphysik, Technische Universit\"{a}t Darmstadt, D-64289 Darmstadt,
Germany}

\author{A. \surname{Richter}}
\affiliation{Institut f\"{u}r Kernphysik, Technische Universit\"{a}t
Darmstadt, D-64289 Darmstadt, Germany} \affiliation{ECT*, Villa
Tambosi, I-38123 Villazzano (Trento), Italy}

\author{C. \surname{Forss\'{e}n}}
\affiliation{Fundamental Physics, Chalmers University of Technology,
SE-412 96 G\"{o}teborg, Sweden}

\author{B. A. \surname{Brown}}
\affiliation{Department of Physics and Astronomy, and National
Superconducting Cyclotron Laboratory, Michigan State University, East
Lansing, Michigan 48824-1321, USA}

\date{\today}

\begin{abstract}

Spectra of the $^{9}$Be($e,e^{\prime}$) reaction have been measured at
the S-DALINAC at an electron energy $E_{0}= 73$~MeV and scattering
angles of $93^{\circ}$ and $141^{\circ}$ with high energy resolution up
to excitation energies $E_{x} = 8$~MeV. The astrophysically relevant
resonance parameters of the first excited $1/2^{+}$ state of $^{9}$Be
have been extracted in a one-level approximation of $R$-matrix theory
resulting in a resonance energy $E_{R} = 1.748(6)$~MeV and width
$\Gamma_{R} = 274(8)$~keV in good agreement with the latest
$^{9}$Be($\gamma,n$) experiment but with considerably improved
uncertainties. However, the reduced $B(E1)$ transition strength deduced
from an extrapolation of the $(e,e')$ data to the photon point is a
factor of two smaller. Implications of the new results for a possible
production of $^{12}$C in neutron-rich astrophysical scenarios are
discussed.

\end{abstract}


\maketitle

\section{Introduction}

The nucleus $^{9}$Be is a loosely-bound system formed by two $\alpha$
particles and a neutron where none of any two constituents alone can
form a bound system. It has the lowest neutron threshold S$_{n}$ =
1.6654 MeV of all stable nuclei. Already the first excited state lies
some tens of keV above the neutron threshold and thus all excited
states are unstable with respect to neutron decay.

The properties of the first excited state are of particular interest
because they determine the importance of $^9$Be production for the
synthesis of $^{12}$C seed material triggering the $r$ process in type
II supernovae~\cite{woosley1992,meyer1992,howard1993,woosley1994}. In
stellar burning the triple $\alpha$ process dominates the production of
$^{12}$C, where at sufficiently high temperatures a small equilibrium
amount of the short-lived $^{8}$Be is formed, which can capture the
third $\alpha$ particle and form $^{12}$C. But in explosive
nucleosynthesis, such as a core-collapse supernova, the reaction path
$^{8}$Be$(n,\gamma)^{9}$Be$(\alpha,n)^{12}$C may provide an alternative
route for building up heavy elements.

A direct measurement of the cross sections of the
$^{8}$Be$(n,\gamma)^{9}$Be reaction is impossible because of the short
life time of about $10^{-16}$~s of the $^{8}$Be ground state (g.s.) but
they can be deduced from  photodisintegration cross sections on
$^{9}$Be using the principle of detailed balance. At low energies, the
photodisintegration cross section is dominated by the properties of the
$1/2^{+}$ resonance just above the $^{8}{\rm Be} +n$ threshold in
$^{9}$Be. The description of this unbound level, viz.\ its resonance
energy and width is a long-standing problem. Due to its closeness to
the neutron threshold the resonance has a strongly asymmetric line
shape.

Several experiments have investigated the $^{9}$Be($\gamma,n)$
reaction, either with real photons from bremsstrahlung or from
laser-induced Compton backscattering, and with virtual photons from
electron scattering (see Ref.~\cite{tilley2004} for a discussion and
references). But despite the sizable body of data there still exist
considerable uncertainties of the resonance parameters.
\textcite{utsunomiya2000} measured the photoneutron cross section for
$^{9}$Be with real photons in the whole energy range of astrophysical
relevance. The deduced resonance parameters for the $1/2^{+}$ state are
shown in Tab.~\ref{tab:FExStExpResPar} in comparison with results from
earlier electron scattering
experiments~\cite{kuechler1987,glickman1991}. Another recent result
from $^9$Li $\beta$ decay ($E_R = 1.689(10)$~MeV, $\Gamma_R =
224(7)$~keV) is quoted by Ref.~\cite{mukha2005} but it is not clear
whether these values refer to the true resonance parameters or the peak
and and full width at half maximum (FWHM). Furthermore, the results
have been questioned by \textcite{barker2006}. Table~\ref{tab:FExStExpResPar}
provides a summary of resonance parameters deduced from the various
experiments.
%\squeezetable
\begin{table}[tbh]
\caption{Summary of resonance parameters and reduced transition probability of
the $1/2^{+}$ state in $^{9}$Be deduced from different experiments.
Reference~\cite{barker2000} contains a reanalysis of the data
of~\cite{kuechler1987}.}
    \begin{center}
    \begin{tabular}{cclll}
    \hline\hline
    Reaction & Ref. & $E_R$ (MeV) & $\Gamma_R$ (keV) & $B(E1)\!\!\uparrow$ ($e^2$fm$^2$) \\
    \hline
    (e,e$^\prime$) & \cite{kuechler1987}   & 1.684(7)  &  217(10) & 0.027(2)  \\
    (e,e$^\prime$) & \cite{glickman1991}   & 1.68(15)  &  200(20) & 0.034(3)  \\
    ($\gamma,n$)    & \cite{utsunomiya2000} & 1.750(10)   &  283(42) & 0.0535(35)\\
    (e,e$^\prime$) & \cite{barker2000}     & 1.732     &  270     & 0.0685    \\
    $\beta$-decay   & \cite{mukha2005}      & 1.689(10) &  224(7)  &   -       \\
    (e,e$^\prime$) & present               & 1.748(6)  &  274(8)  & 0.027(2)  \\
    \hline\hline
    \end{tabular}
    \label{tab:FExStExpResPar}
    \end{center}
\end{table}

Obviously there are significant differences between the results obtained
from photonuclear and electron scattering
experiments. The discrepancy in the $B(E1)$ transition strength amounts
to a factor of about two when comparing with the results of
($e,e^{\prime}$) measurements~\cite{kuechler1987,clerc1968} at low
momentum transfers $q$, while it is reduced by $\sim$~30\% in an
($e,e^{\prime}$) experiment~\cite{glickman1991} at larger $q$. The
reason for these discrepancies between the strengths deduced from the
real photon and virtual photon experiments is unknown. Furthermore,
\textcite{barker2000} reanalyzed the ($e,e^{\prime}$) data of
\textcite{kuechler1987} and extracted resonance parameters which differ
considerably from those quoted in the original paper (see
Tab.~\ref{tab:FExStExpResPar}).

In order to resolve these discrepancies, high-resolution measurements
of the  $^{9}$Be($e,e^{\prime}$) reaction were performed at the
S-DALINAC and new resonance parameters for the $1/2^+$ state are
extracted. Furthermore, an independent reanalysis of the electron
scattering data of \textcite{kuechler1987} was performed. Finally, the
temperature dependence of the $^{8}$Be$(n,\gamma)^{9}$Be reaction rate
including the new results is derived and compared to that of the NACRE
compilation \cite{NACRE1999} serving as standard for most network
calculations.

\section{Experiment}

The $^{9}$Be($e,e^{\prime}$) experiment was carried out at the
high-resolution 169$^{\circ}$ magnetic spectrometer of the
Superconducting Darmstadt Electron Linear Accelerator
(\mbox{S-DALINAC}). Data were taken at an incident electron beam energy
$E_{0} = 73$~MeV and scattering angles $\Theta_{Lab} = 93^{\circ}$ and
141$^{\circ}$ with typical beam currents of 2~$\mu$A. For the
measurements a self-supporting $^{9}$Be target with an areal density of
5.55 mg/cm$^{2}$ was used. The properties of the spectrometer are
described in Ref.~\cite{walcher1978}. A new focal plane detector system
based on silicon microstrip detectors was recently
implemented~\cite{lenhardt2006}. In dispersion-matching mode an energy
resolution $\Delta E \simeq 30$~keV (full width at half maximum, FWHM)
was achieved in the measurements.

\begin{figure}[h]
\includegraphics[angle=0,width=8.5cm]{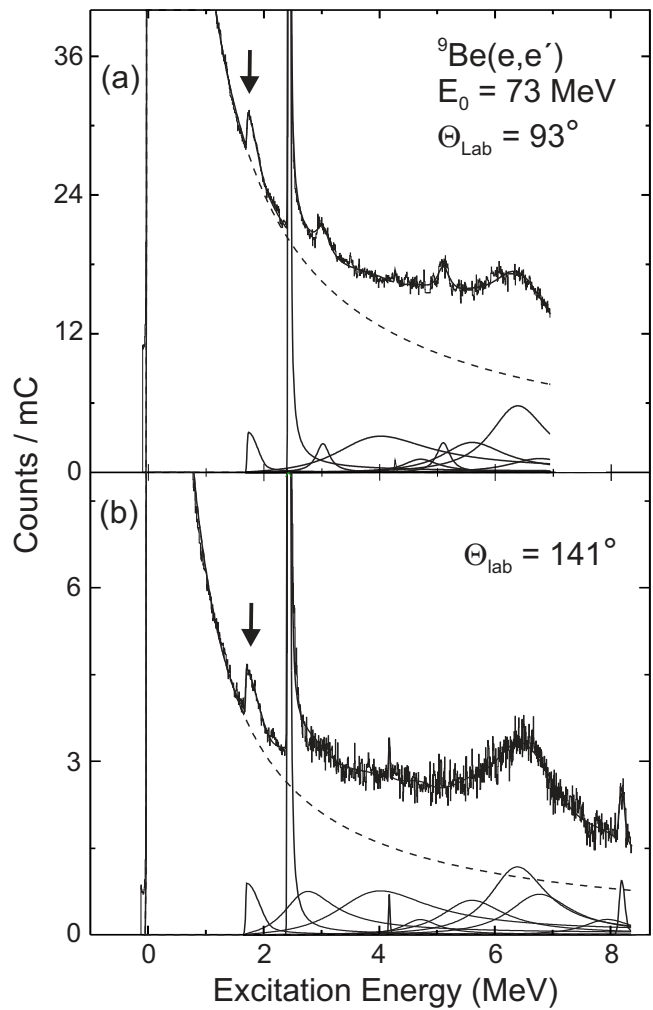}
\caption{\label{fig:ExpSpe} Spectra of the $^{9}$Be($e,e^{\prime}$)
reaction at $E_{0}=73$ MeV and $\Theta_{lab}= 93^{\circ}$ (top) and
$141^{\circ}$ (bottom) and their decomposition. Solid lines: Fits to
experimentally known resonances. Dashed lines: Radiative tail from
elastic scattering. The arrows indicate the transition to the first excited state,
whose asymmetric line shape is clearly visible.}
\end{figure}
Figure~\ref{fig:ExpSpe} presents the spectra of the
$^{9}$Be($e,e^{\prime}$) reaction measured up to an excitation energy
of about 8 MeV. There is only one narrow peak visible in the spectra
which corresponds to the excitation of the $J^\pi = 5/2^{-}$ state at
$E_{x} =2.429$~MeV. The tiny peak at about 4.2~MeV in both spectra
correspond to the first excited state in $^{12}$C (the deviation
between the observed and the true excitation energy $E_{x} = 4.439$~MeV
stems from the difference of the recoil correction for nuclei with mass
$A = 9$ and 12, respectively). The bumps around 5.1~MeV in the top
spectrum and around 8.1~MeV in the bottom spectrum are due to elastic
scattering on hydrogen. The broad bump between 6 and 7~MeV results from
the overlap of resonances at $E_{x} =6.38$~MeV ($J^{\pi} = 7/2^{-}$)
and $E_{x} = 6.76$~MeV ($J^{\pi} = 9/2^{+}$) in
$^{9}$Be~\cite{glickman1991}. The asymmetric line shape of the
$1/2^{+}$ state at $E_{x} \approx 1.7$~MeV (marked by arrows
Fig.~\ref{fig:ExpSpe}) in is already clearly visible in the raw
spectra.

In the decomposition of the spectra, the line shape of the narrow state
was described by the function given in Ref.~\cite{hofmann2002}, whose
parameters were determined by a fit to the elastic line. This also
determines the background from the radiative tail of the elastic line
in the region of interest indicated by the dashed lines in
Fig.~\ref{fig:ExpSpe}. The line shapes of the broad resonances were
assumed to correspond to an energy-dependent Breit-Wigner function
discussed below. The energies and widths of the resonances (taken from
the latest compilation~\cite{tilley2004}) were kept fixed during the
fit except for the parameters of the first excited state. Finally, the
$1/2^+$ resonance was treated in a one-level $R$-matrix formalism
explained in the next section.

\section{Analysis}

Since the state of interest lies above the neutron threshold, we first
discuss an extraction of the relevant parameters as if it was excited
in a $(\gamma,n)$ reaction. In Sec.~\ref{sec:be1}, the relation to the
$(e,e')$ data is explained. Finally, Sec.~\ref{sec:resonance} describes
the extraction of the resonance parameters.

\subsection{One-level $R$-matrix approximation}
\label{sec:r-matrix}

The contribution to the $(\gamma,n)$ cross section from an isolated
level of spin $J$ located near threshold  in the one-level
approximation of $R$-matrix theory~\cite{lane1958} is given by
\begin{equation}
\sigma _{\gamma,n} (E_{\gamma} ) = \frac{\pi}{{2\,k_{\gamma}^2 }} \, \frac{2J + 1}{2I +
1} \, \frac{{\Gamma _{\gamma} \, \Gamma _{n} }}{{\left( {E_{\gamma} - E_{\lambda} -
\Delta(E)}\right)^2 + \frac{{\displaystyle \Gamma ^2}}{{\displaystyle 4}}}},
\label{eq:CSgnCom}
\end{equation}
where $k_{\gamma} = E_{\gamma} / \hbar c$ stands for the photon wave
number, $I$ for ground state spin, $\Gamma_{\gamma}$ for the ground
state radiative width, $\Gamma_{n}$ for the neutron decay width, the
total decay width $\Gamma = \Gamma_{\gamma} + \Gamma_{n}$, and
$E_{\lambda}$ corresponds to the energy eigenvalue. The level shift
$\Delta(E)$ is given by
\begin{equation}
\Delta(E) = -\gamma^{2}\,(S(E) - B), \label{eq:EnergyShiftDelta}
\end{equation}
with the reduced width $\gamma^{2}$, the shift factor $S(E)$ and the
boundary condition parameter $B$ (see Ref.~\cite{lane1958}).

Then for a $1/2^{+}$ level in $^{9}$Be excited by $E1$ $\gamma$
radiation and decaying by $s$-wave neutrons, and for an energy $E =
E_{\gamma} -S_{n} > 0$, one has
\begin{equation}
\Gamma_{\gamma} = \frac{16 \, \pi}{9} \, e^{2} \, k^{3}_{\gamma} \,
B(E1,k)\!\!\downarrow,
\end{equation}
\begin{equation}
\Gamma_{n} = 2 \, \sqrt{\epsilon \, (E_{\gamma} - S_{n})}, \label{eq:NeutronWidth}
\end{equation}
%%
%\begin{equation}
%\Delta(E) = 0,
%\end{equation}
%%
with $k_\gamma = E_x/\hbar c$ being the photon momentum transfer
(called photon point), $B(E1,k)\!\!\downarrow$ the reduced transition
strength at the photon point for the decay, $\epsilon = 2 \mu a^2
\gamma^4 / \hbar^2 > 0$, where $\mu$ and $a$ are the reduced mass and
the $^{8}{\rm Be}+ n$ channel radius, respectively, and $S_{n}(^{9}{\rm
Be}) = 1.6654$~MeV \cite{tilley2004} the neutron threshold energy. The
boundary condition parameter $B$ is taken to be zero and the shift
factor $S(E)\,=\,0$ for $s$-wave neutrons~\cite{lane1958} and thus,
$\Delta(E) = 0$.

Since $\Gamma_{n} \gg \Gamma_{\gamma}$, the total resonance width
$\Gamma \approx \Gamma_{n}$ and the energy dependence of the
photoabsorption cross section of Eq.~(\ref{eq:CSgnCom}) is reduced to
\begin{equation}
\begin{array}{cl}
\sigma _{\gamma,n} (E_{\gamma} ) = & {\displaystyle \frac{16 \, \pi^{2}}{9} \,
\frac{e^{2}}{\hbar c} \,
\frac{2J + 1}{2I + 1} \, B(E1,k)\!\!\downarrow} \\
& {\displaystyle \times \, \frac{{E_{\gamma} \, \sqrt{\epsilon\,(E_{\gamma} -
S_{n})}}}{{\left( {E_{\gamma} - E_{R} } \right)^2  + \epsilon\,(E_{\gamma} - S_{n})}}}.
\label{eq:CSgn}
\end{array}
\end{equation}
The resonance energy $E_{R}$ is calculated from
\begin{equation}
E_{R} = E_{\lambda} + \Delta,
\end{equation}
and the resonance width using Eq.~(\ref{eq:NeutronWidth})
\begin{equation}
\Gamma_{R}(E_{R})= 2 \, \sqrt{\epsilon(E_{R} - S_{n})}.
\label{eq:ResonanceEnergy}
\end{equation}
It should be noted that because of the asymmetric line shape the
resonance energy $E_{R}$ does not coincide with the excitation energy
at the maximum of the cross section, and the resonance width
$\Gamma_{R}$ differs from the FWHM.

\subsection{Extraction of equivalent $(\gamma,n)$ cross sections and
$B(E1)$ transition strength from the $(e,e')$ data} \label{sec:be1}

Equation (\ref{eq:CSgn}) holds also for the relation between the
$(e,e')$ cross sections and the reduced transition strength if
$B(E1,k)$ is replaced by the corresponding value at finite momentum
transfer $B(E1,q)$. (Note that in the following  $B(E1,q)\!\!\uparrow =
(2J_f+1)/(2J_i+1) B(E1,q)\!\!\downarrow$ is given, where $J_{i,f}$
denote the spins of initial and final state, respectively). If
interference with transitions to higher-lying $1/2^+$ resonances can be
neglected, the equivalent $\sigma_{\gamma,n}$ cross sections can be
determined from the electron scattering results by extrapolating the
reduced transition strength $B(E1,q)$ measured at finite momentum
transfer $q$ to the photon point $k = E_{x}/\hbar c$.

\begin{figure}[h]
\includegraphics[angle=0,width=8.5cm]{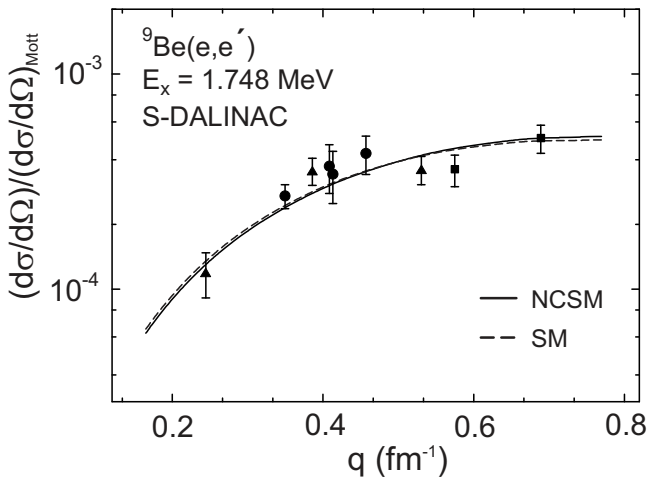}
\caption{Ratio of the measured cross sections to the Mott cross section
of the transition to the $1/2^{+}$ state in $^{9}$Be as a
function of momentum transfer. Data are from
Ref.~\cite{clerc1968} (triangles), Ref.~\cite{kuechler1987}
(circles) and present work (squares). Solid and dashed lines are
theoretical predictions of the SM and NCSM calculations
described in the text normalized to the data.}
\label{fig:eeFFqNCSMplSM}
\end{figure}
Figure~\ref{fig:eeFFqNCSMplSM} presents the momentum transfer
dependence of the measured $(e,e^{\prime})$ cross sections normalized
to the Mott cross section for the transition to the first $1/2^{+}$
state in $^{9}$Be. Besides the data from the present work displayed as
squares, results of previous experiments at comparable momentum
transfers shown as triangles (Ref.~\cite{clerc1968}) and circles
(Ref.~\cite{kuechler1987}) are included. In first-order perturbation
theory inclusive electron scattering cross sections factorize in a
longitudinal ($C$) and a transverse ($E$) part reflecting the
respective polarization of the exchanged virtual photon. The kinematics
of the data shown in Fig.~\ref{fig:eeFFqNCSMplSM} favor longitudinal
excitation and thus $B(C1,q)$ rather than $B(E1,q)$ is determined. Both
quantities can be related by Siegert$^{\prime}$s theorem
$B(E1,q)$=$(k/q)^{2}\,B(C1,q)$, i.e., they should be equal at the
photon point $q = k$.

There are two methods to perform the extrapolation from finite momentum
transfer to the photon point: (i) based on microscopic model
calculations, or (ii) the plane wave Born approximation (PWBA) for a
nearly model-independent extraction. The latter method is valid only at
small momentum transfers ($q < 1$~fm$^{-1}$) and for small atomic
numbers $Z$ ($\alpha Z \ll 1$).

For an application of the first method shell-model (SM) calculations of
the electroexcitation of the $1/2^+$ state were performed with the
interaction of \textcite{warburton1992} coupling $1p$ and
$2s1d$-shells. The formalism for calculating electron scattering form
factors from the shell-model one-body transition densities is described
in Ref.~\cite{brown1985}. A similar calculation of an $E1$ longitudinal
form factor for a transition in $^{12}$C is described in
Ref.~\cite{campos1995}. Spurious states are removed with the
Gloeckner-Lawson method \cite{gloeckner1974}. The SM one-body
transition density is dominated by the $0p_{1/2} \rightarrow 1s_{1/2}$
neutron transition. For these two orbitals we used Hartree-Fock radial
wave functions obtained with the SKX Skyrme interaction
\cite{brown1998} with their separation energies constrained to be 1.665
MeV and 0.2 MeV, respectively. Harmonic oscillator radial wave
functions were used for all other orbitals. The result normalized to
the data is shown in Fig.~\ref{fig:eeFFqNCSMplSM} as dashed line.

Alternatively, a no-core shell model calculation (NCSM) was performed
in the framework of the model described in Ref.~\cite{forssen2005}
(solid line in Fig.~\ref{fig:eeFFqNCSMplSM}). This calculation utilized the realistic nucleon-nucleon interaction CD-Bonn 2000 using very large model spaces, viz.\ 8(9) $\hbar\omega$ for the $3/2^- (1/2^+)$ state,
and a HO frequency of 12 MeV. Despite the large model spaces and
improved convergence techniques \cite{forssen2008}, no convergence was
achieved for the wave function of the $^9$Be, $1/2^+$ state. One should note that these calculations treat the $1/2^+$ state in a quasibound
approximation.

The two calculations predict a very similar momentum transfer
dependence which describes the data well. However, the absolute
magnitudes are underpredicted by factors 3.6 (SM) and 1.7 (NCSM),
respectively. Normalizing the theoretical predictions
($B(E1,k)\!\!\uparrow = 0.008$~$e^2$fm$^2$ (SM) and 0.016~$e^2$fm$^2$
(NCSM), respectively) to the experimental data one finds
$B(E1,k)\!\!\uparrow$~= 0.027(2)~$e^{2}$fm$^{2}$ using the NCSM and
$B(E1,k)\!\!\uparrow$~= 0.029(2)~$e^{2}$fm$^{2}$ using the SM form
factor. Both results agree with each other within error bars.

An alternative independent method to derive the E1 transition strength
is based on a PWBA analysis (see e.g.\ Ref.~\cite{theissen1972}). At
low momentum transfers the form factor can be expanded in a power
series of $q$
\begin{equation}
\label{eq:pwba}
\sqrt{B(E1,q)} = \sqrt{B(E1,0)} \left(1 - \frac{R_{tr}^2 q^2}{10}
+ \frac{R_{tr}^4 q^4}{280} - \dots \right) \ ,
\end{equation}
where higher powers of $q$ are negligible in the momentum transfer
range studied in the present experiment. The so-called transition
radius $R_{tr}$ is given by $R_{tr}^2 = \langle r^{\lambda + 2}
\rangle_{tr} / \langle r^{\lambda} \rangle_{tr}$, where $\langle
r^{\lambda} \rangle_{tr}$ denotes the moments of the transition density
\begin{equation}
\label{eq:trd}
\langle r^{\lambda} \rangle_{tr} = 4 \pi \int \rho_{tr} r^{\lambda+2} dr \ .
\end{equation}
An additional assumption is made that $R_{tr}^4$ can be parameterized
in the form $R_{tr}^4 = a (R_{tr}^2)^2$, where the parameter $a$ is
determined using theoretical transition densities.

\begin{figure}[h]
\includegraphics[angle=0,width=8.5cm]{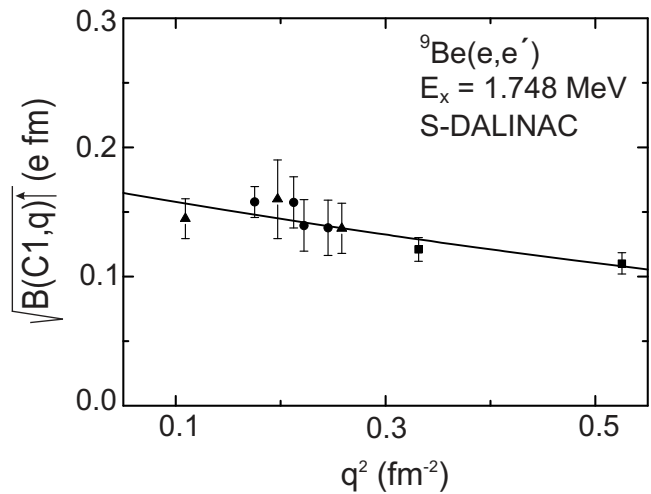}
\caption{Ratio of the measured cross sections of the transition to
the $1/2^{+}$ state in $^{9}$Be to the Mott cross sections as a
function of the squared momentum transfer. The solid line is a fit
of Eq.~(\ref{eq:pwba}) with parameters $\sqrt{B(C1,0)} = 0.164(12)$~$e$fm and $R_{tr} =2.9(3)$~fm} \label{fig:eePWBAExtrapToPhPoint}
\end{figure}
Since the above relation (\ref{eq:pwba}) holds in the plane wave limit
only, distorted wave Born approximation (DWBA) correction factors have
been calculated based on the NCSM results in order to convert the
measured cross sections into equivalent PWBA cross sections.
Corrections of the order 10\% are obtained.
Figure~\ref{fig:eePWBAExtrapToPhPoint} presents the so corrected data
as a function of the squared momentum transfer. The solid line shows a
fit of Eq.~(\ref{eq:pwba}) with parameters $\sqrt{B(C1,0)}=
0.164(12)$~$e$fm and $R_{tr} =2.9(3)$~fm. Extrapolation
of the transition strength to the photon point yields
$B(E1,k)\!\!\uparrow = 0.027(4)$~$e^{2}$fm$^{2}$ in agreement with the
results obtained from the analysis based on microscopic form factors.

A significant difference to the corresponding  $B(E1,k)\!\!\uparrow$
strength deduced from the real-photon experiment is observed, which
finds 0.0535(35)~e$^2$fm$^2$ (cf. Tab.~\ref{tab:FExStExpResPar}), about
a factor of two larger than the present result. This implies a severe
violation of Siegert's theorem. Its origin is presently unclear but
possible explanations could lie in the quasibound approximation used in
the shell-model calculations and/or a need to modify the $E1$ operator.
A detailed discussion of this interesting problem is postponed to a
future publication.

\subsection{Resonance parameters}
\label{sec:resonance}

\begin{figure}[h]
\includegraphics[angle=0,width=8.5cm]{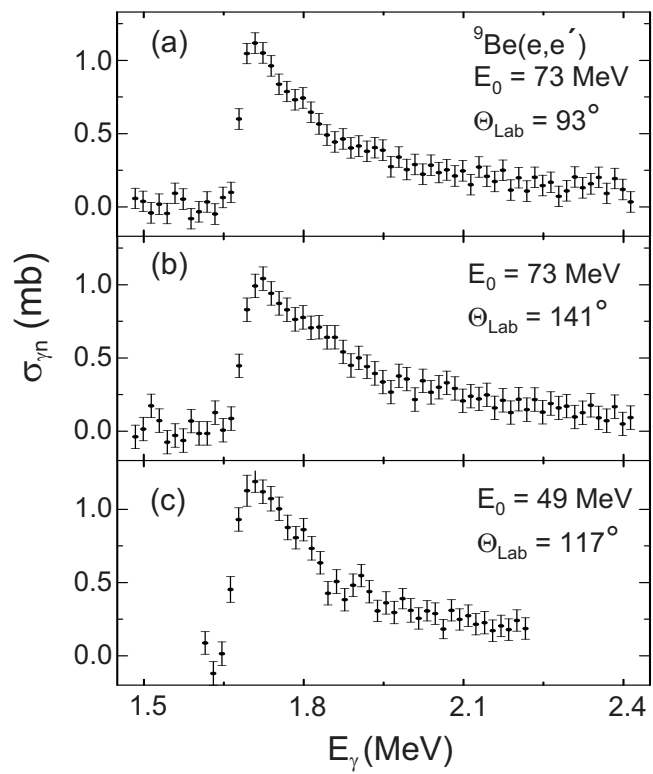}
\caption{Photoneutron cross sections extracted from the present (top and middle) and
older (bottom) ($e,e^{\prime}$) data~\cite{kuechler1987}.}
\label{fig:SgnAll}
\end{figure}

Figure~\ref{fig:SgnAll} shows the photoneutron cross sections of the
first excited state in $^9$Be extracted from the present work (top and
middle) together with the previous (bottom) result of
Ref.~\cite{kuechler1987}. The data are summed in 15 keV bins. All three
data sets are in good agreement with each other.

\begin{figure}[h]
\includegraphics[angle=0,width=8.5cm]{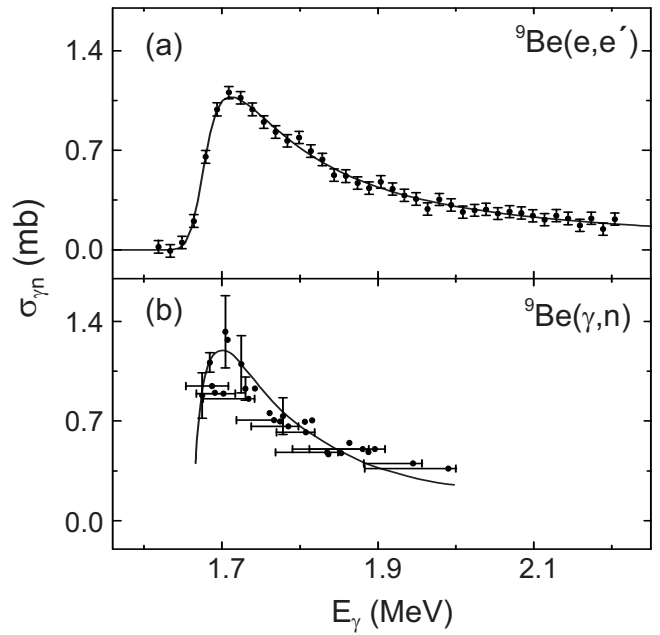}
\caption{Averaged photoneutron cross sections extracted from the ($e,e^{\prime}$) data
shown in Fig.~\ref{fig:SgnAll} in comparison with the cross sections extracted from the
latest $^9$Be$(\gamma,n)$ experiments~\cite{utsunomiya2000}. The solid lines are the
corresponding fits with Eq.~(\ref{eq:CSgn}) with the parameters given in the text.}
\label{fig:SgnRealVirtComp}
\end{figure}

Since all three measurements shown in Fig.~\ref{fig:SgnAll} were
independent, the data can be averaged. The resulting averaged
($\gamma,n$) cross sections are presented in the upper part of
Fig.~\ref{fig:SgnRealVirtComp}.
The solid line is a fit with Eq.~(\ref{eq:CSgn}). In order to account
for the detector response the theoretical form is folded with the
experimental resolution function. Since the experimental resolution was
much smaller than the resonance width, the influence of the resolution
function is small except for energies around the maximum of the cross
sections. The fit results in a resonance energy $E_{R} = 1.748(6)$~MeV
and a width $\Gamma_{R} = 274(8)$~keV in contradiction to the results
of Ref.~\cite{kuechler1987} but in agreement with the reanalysis of
\textcite{barker2000}. In fact, since the data of
\textcite{kuechler1987} are very close to those of the present work
(cf.\ Fig.~\ref{fig:SgnAll}) an independent reanalysis yields resonance
parameters very similar to the ones from the new data. The most likely
explanation for the values given in Ref.~\cite{kuechler1987} is that
the maximum energy and FWHM instead of the true resonance parameters
were quoted. The final results are included in Tab.~\ref{tab:FExStExpResPar}.

The lower part of Fig.~\ref{fig:SgnRealVirtComp} shows the measured
$^9$Be$(\gamma,n)$ cross sections of \textcite{utsunomiya2000}.
Application of Eq.~(\ref{eq:CSgn} leads to comparable resonance
parameters $E_{R} = 1.750(10)$~MeV and $\Gamma_{R} = 283(42)$~keV but
the present work provides values with considerably improved
uncertainties.

\section{Astrophysical implications}

To calculate the thermonuclear reaction rate of $\alpha(\alpha
n,\gamma)^{9}$Be in a wide range of temperatures, we numerically
integrate the thermal average of cross sections $N^{2}_{A}\langle\sigma
v\rangle$  (as defined e.g.\ in Ref.~\cite{NACRE1999}) assuming
two-step formation of $^{9}$Be through a metastable $^{8}$Be. The
formation through $^{5}$He followed by an $\alpha$ capture is generally
neglected because of the short lifetime of $^{5}$He except for the work
of Ref.~\cite{buchmann2001} which indicates relevance of this channel
at $T \gg T_9$ (see, however, the criticism in
Ref.~\cite{grigorenko2005}). The same formulation to the $^{9}$Be
formation is also used in the NACRE compilation~\cite{NACRE1999}.
Resonant and non-resonant contributions from the $\alpha + \alpha
\rightarrow ^{8}$Be reaction are taken into account. The ground state
of $^8$Be is described by a resonance energy $E_{R} = 0.0918$~MeV with
respect to $\alpha + \alpha$ threshold and  a width $\Gamma_{\alpha} =
5.57(25)$~eV taken from Ref.~\cite{tilley2004}. Elastic cross sections
of $\alpha \alpha$ scattering were treated as described in
Ref.~\cite{nomoto1985}.

The resonance properties (energy, $\gamma$ and neutron decay widths) of
the lowest excited states in $^{9}$Be with the corresponding g.s.\
branching ratios $f$ included into the calculation of the
$\alpha(\alpha n,\gamma)^{9}$Be reaction rate are summarized in
Tab.~\ref{tab:StContrToRate}. An energy dependence of the partial decay
widths was taken into account only for the $1/2^{+}$ resonance.
%
%\squeezetable
\begin{table}[tbh]
\caption{Low-lying states in $^{9}$Be considered in the calculations of
the $\alpha(\alpha n,\gamma)^{9}$Be reaction rate. The quantity $f$
denotes the branching ratio of the corresponding state into the
$n\,+\,^{8}$Be decay channel.}
    \begin{center}
    \begin{tabular}{cccccc}
    \hline\hline
$J^{\pi}$ & $E_{R}$ (MeV) & $\Gamma_{\gamma}$ (eV) & $\Gamma_{n}$ (MeV) & $f$ $(\%)$ & Ref.\\
    \hline
$1/2^{+}$ & 1.748(6) & 0.302(45) & 0.274(8) & 100 & Present\\
$5/2^{-}$ & 2.4294(13) & 0.089(10) & 0.78(13)  & 7(1) & \cite{tilley2004}\\
$1/2^{-}$ & 2.78(12)  & 0.45(36)  & 1.08(11)  & 100 & \cite{NACRE1999,tilley2004}\\
$5/2^{+}$ & 3.049(9) & 0.90(45)   & 0.282(110) & 87(13) & \cite{NACRE1999,tilley2004}\\
    \hline\hline
    \end{tabular}
    \label{tab:StContrToRate}
    \end{center}
\end{table}

\begin{figure}[h]
\includegraphics[angle=0,width=8.5cm]{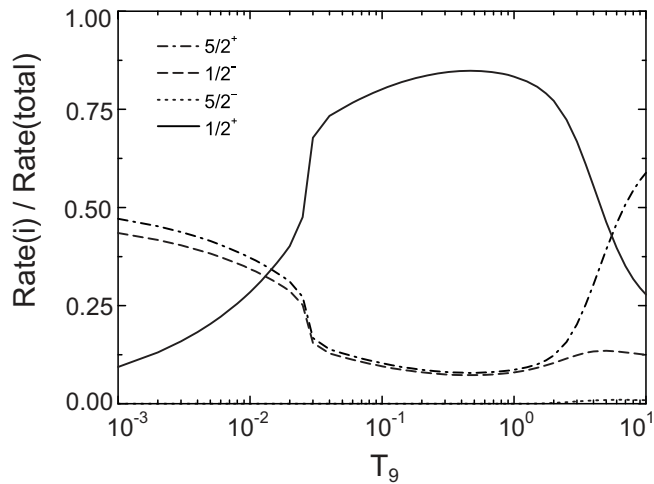}
\caption{Contributions of the lowest-lying states (i) in $^{9}$Be to the
$\alpha(\alpha n,\gamma)^{9}$Be reaction rate.}
\label{fig:ReactRateContr}
\end{figure}
Reaction rates calculated at representative temperatures are given in
Tab.~\ref{tab:ReactRate}.
%
%\squeezetable
\begin{table}[tbh]
\squeezetable
    \caption{The thermonuclear reaction rate $N^{2}_{A}\langle\sigma v\rangle$ of
$\alpha(\alpha n,\gamma)^{9}$Be at representative temperatures.}
    \begin{center}
    \begin{tabular}{lc|lc|lc}
    \hline\hline
$T_{9}$ & Rate & $T_{9}$ & Rate & $T_{9}$ & Rate \\
    \hline
0.001 & 4.67$\cdot10^{-59}$ & 0.04  & 7.53$\cdot10^{-16}$ & 0.5  & 3.93$\cdot10^{-07}$ \\
0.002 & 2.82$\cdot10^{-47}$ & 0.05  & 1.07$\cdot10^{-13}$ & 0.6  & 3.91$\cdot10^{-07}$ \\
0.003 & 1.45$\cdot10^{-41}$ & 0.06  & 2.74$\cdot10^{-12}$ & 0.7  & 3.70$\cdot10^{-07}$ \\
0.004 & 5.77$\cdot10^{-38}$ & 0.07  & 2.68$\cdot10^{-11}$ & 0.8  & 3.41$\cdot10^{-07}$ \\
0.005 & 2.11$\cdot10^{-35}$ & 0.08  & 1.43$\cdot10^{-10}$ & 0.9  & 3.11$\cdot10^{-07}$ \\
0.006 & 1.90$\cdot10^{-33}$ & 0.09  & 5.17$\cdot10^{-10}$ & 1    & 2.81$\cdot10^{-07}$ \\
0.007 & 6.97$\cdot10^{-32}$ & 0.1   & 1.41$\cdot10^{-09}$ & 1.25 & 2.18$\cdot10^{-07}$ \\
0.008 & 1.36$\cdot10^{-30}$ & 0.11  & 3.17$\cdot10^{-09}$ & 1.5  & 1.71$\cdot10^{-07}$ \\
0.009 & 1.69$\cdot10^{-29}$ & 0.12  & 6.12$\cdot10^{-09}$ & 1.75 & 1.37$\cdot10^{-07}$ \\
0.01  & 1.49$\cdot10^{-28}$ & 0.13  & 1.06$\cdot10^{-08}$ & 2    & 1.12$\cdot10^{-07}$ \\
0.011 & 9.96$\cdot10^{-28}$ & 0.14  & 1.67$\cdot10^{-08}$ & 2.5  & 7.90$\cdot10^{-08}$ \\
0.012 & 5.38$\cdot10^{-27}$ & 0.15  & 2.46$\cdot10^{-08}$ & 3    & 6.00$\cdot10^{-08}$ \\
0.013 & 2.44$\cdot10^{-26}$ & 0.16  & 3.43$\cdot10^{-08}$ & 3.5  & 4.81$\cdot10^{-08}$ \\
0.014 & 9.60$\cdot10^{-26}$ & 0.18  & 5.86$\cdot10^{-08}$ & 4    & 4.00$\cdot10^{-08}$ \\
0.015 & 3.34$\cdot10^{-25}$ & 0.2   & 8.79$\cdot10^{-08}$ & 5    & 2.97$\cdot10^{-08}$ \\
0.016 & 1.05$\cdot10^{-24}$ & 0.25  & 1.71$\cdot10^{-07}$ & 6    & 2.32$\cdot10^{-08}$ \\
0.018 & 7.98$\cdot10^{-24}$ & 0.3   & 2.50$\cdot10^{-07}$ & 7    & 1.87$\cdot10^{-08}$ \\
0.02  & 4.65$\cdot10^{-23}$ & 0.35  & 3.11$\cdot10^{-07}$ & 8    & 1.53$\cdot10^{-08}$ \\
0.025 & 1.86$\cdot10^{-21}$ & 0.4   & 3.54$\cdot10^{-07}$ & 9    & 1.27$\cdot10^{-08}$ \\
0.03  & 1.97$\cdot10^{-19}$ & 0.45  & 3.80$\cdot10^{-07}$ & 10   & 1.06$\cdot10^{-08}$ \\
   \hline\hline
    \end{tabular}
    \label{tab:ReactRate}
    \end{center}
\end{table}

Figure~\ref{fig:ReactRateContr} shows the individual contributions of
the excited states considered in Tab.~\ref{tab:StContrToRate} to the
total reaction rate as a function of temperature. The $1/2^{+}$ state
(solid line) dominates in the temperature range $T_9 = 0.04 - 3$. The
role of the $5/2^{-}$ state (dotted line) is negligible small at all
temperatures. At values $T_9 < 0.04$ the low-energy tails of the broad
$1/2^{-}$ (dashed line) and $5/2^{+}$ (dashed-dotted line) resonances
become increasingly important. Temperatures in supernova II scenarios
reach values well above $T_9$. Under these conditions, the maximum of
the photon spectrum is shifted to energies above the $1/2^+$ state and
the $5/2^+$ state starts to dominate when approaching $T_9 = 10$.

\begin{figure}[h]
\includegraphics[angle=0,width=8.5cm]{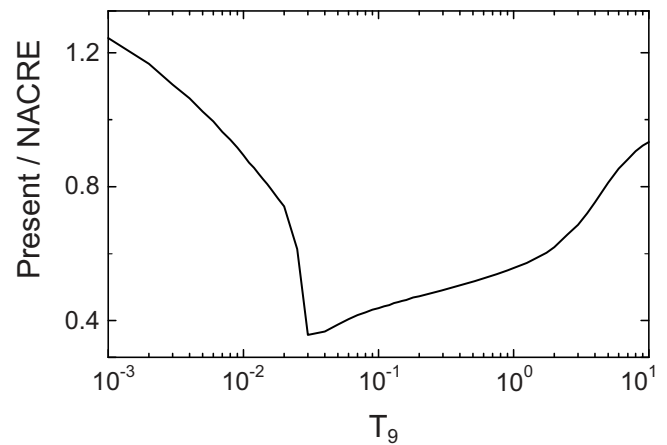}
\caption{The ratio of the present rate to the latest NACRE compilation~\cite{NACRE1999}.
Deviations ranges from +25$\%$ to -60$\%$ depending on the temperature.}
\label{fig:RatioPresentToNacre}
\end{figure}
The ratio of the present reaction rates to the latest NACRE
compilation~\cite{NACRE1999} is shown in Fig.~\ref{fig:RatioPresentToNacre}. Deviations ranges from +20$\%$ to
-60$\%$ depending on the temperature. Besides using the improved resonance parameters of the $1/2^+$ state, there are some differences
between the present calculation and the one described in
Ref.~\cite{NACRE1999}. The $5/2^-$ state is neglected in the latter
case. However, as can be seen in  Fig.~\ref{fig:ReactRateContr} its
contributions are very small. Also the $^8$Be g.s.\ parameters taken
from \cite{tilley2004} differ from those used in Ref.~\cite{NACRE1999}.
The pronounced kink at $T_9 = 0.03$ in
Fig.~\ref{fig:RatioPresentToNacre} marks the onset of resonant
contributions in the $\alpha + \alpha \rightarrow ^{8}$Be cross
sections. Rates from a semimicroscopic three-body
model~\cite{efros1998} are also available for temperatures $0.2 \leq
T_9 \leq 5$. These are typically about 20\% larger than the NACRE
results.

The difference observed for the $\gamma$ decay width of the $1/2^+$
resonance between the measurements of \textcite{utsunomiya2000} and the
present work have a non-negligible impact on the reaction rates. In
general, taking a larger $\Gamma_\gamma$ the contribution of the
$1/2^+$ resonance will increase reducing the deviations from the NACRE
result at high temperatures. It should also be noted that non-resonant
contributions to the $^8$Be$(n,\gamma)$$^9$Be neglected in both
approaches discussed above may be relevant \cite{mengoni1999}. The
calculations described in Refs.~\cite{mengoni1999,goerres1995} suggest
sizable effects while Ref.~\cite{buchmann2001} finds it to be of minor
importance. Finally, there is a recent claim \cite{garrido2010} that
the picture of a sequential formation is incorrect for the
near-threshold $1/2^+$ state and it should be described as a genuine
three-body process \cite{alvarez2008}. This would modify the resonance
parameters considerably.

\section{Concluding remarks}

The astrophysical relevant $^9$Be($\gamma,n$) cross sections have been
extracted from $^9$Be($e,e^\prime$) data. The resonance parameters of
the first excited $1/2^{+}$ state in $^9$Be are derived in a one-level
$R$-matrix approximation. The resonance parameters averaged over all
available ($e,e^{\prime}$) data are $E_{R} = 1.748(6)$~keV and
$\Gamma_{R} = 274(8)$~keV in agreement with the latest direct
($\gamma,n$) experiment \cite{utsunomiya2000} but with much improved
uncertainties. However, the deduced $\gamma$ decay width is about a
factor of two smaller. Rates for the temperature-dependent formation of $^9$Be under stellar conditions are given. They differ significantly
from the values adopted in the NACRE compilation \cite{NACRE1999}.
Further improvements of the reaction rate require the inclusion of  direct capture reactions.

The difference in the $B(E1)$ transition strength obtained from
electron- and photon-induced reaction presents an intriguing problem.
Since the present result is extracted from the longitudinal form
factor, it might indicate a violation of Siegert$^{\prime}$s theorem at the photon point. A similar problem was observed in the
electro-excitation of $1^{-}$ levels in $^{12}$C~\cite{campos1995},
$^{16}$O~\cite{miska1975,friedrich1989}, and
$^{40}$Ca~\cite{graef1977}. There, isospin mixing was offered as an
explanation leading to modified form factors of longitudinal and
transverse electron scattering at small momentum transfers. Another
explanation could be the need for a modification of the $E1$ operator
due to meson-exchange currents. A detailed study of the weak transverse form factor of the transition to the $1/2^+$ resonance in $^9$Be would
be highly desirable to clarify the origin of the discrepancy.

The shell-model calculations seem to describe the momentum transfer
dependence of the electron scattering data for the measured $q$ range
but fall short of the experimental transition strength. One possible
explanation may be the quasibound approximation applied in the
description of the $1/2^+$ state. Near-threshold $\alpha$-cluster
states are expected to have an increased size which amplifies the
dependence on tails of the wave function like e.g.\ observed for the
case of the Hoyle state in $^{12}$C \cite{chernykh2007,chernykh2010}.
Calculations with improved radial wave function would be important.
Also, the role of direct three-body decay needs to be further explored.

\begin{acknowledgments}
H.-D. Gr\"{a}f and the S-DALINAC team is thanked for preparing
excellent beams and M.~Chernykh for his help in data taking. The
experiment originated from discussion of one of us (A.R.) with the late Fred Barker on an inconsistency of the analysis of the data in
Ref.~\cite{kuechler1987}, and we are grateful for Fred's advice.
Discussions with A.~S.~Jensen, G.~Mart\'{\i}nez-Pinedo, A.~Mengoni, and S.~Typel are gratefully acknowledged. This work has been supported by
the DFG under contract SFB 634, and by the NSF, grant PHY-0758099. One
of us (CF) acknowledges financial support from the Swedish Research
Council and the European Research Council under the FP7.
\end{acknowledgments}

% REFERENCES

\bibliography{Be9-v4}

\end{document}